\documentclass[aps,prl,twocolumn,amsmath,amssymb,showpacs,superscriptaddress,notitlepage,longbibliography]{revtex4-1}
\usepackage[colorlinks=true,linkcolor=blue,anchorcolor=red,citecolor=blue, urlcolor=blue]{hyperref}
\usepackage{bm}
\usepackage{graphicx}
\usepackage{color}
\usepackage{cases}
\definecolor{amethyst}{rgb}{0.6, 0.4, 0.8}

\begin{document}

	\title{3D quantum Hall effect in a topological nodal-ring semimetal}

	\author{Guang-Qi Zhao}
	
	\affiliation{Department of Physics, Southern University of Science and Technology (SUSTech),
		Shenzhen 518055, China}
	
	\author{Shuai Li}
	\affiliation{Department of Physics, Southern University of Science and Technology (SUSTech),
		Shenzhen 518055, China}
	\affiliation{Quantum Science Center of Guangdong-Hong Kong-Macao Greater Bay Area (Guangdong), Shenzhen 518045, China}

	\author{W. B. Rui}
	
	\affiliation{Department of Physics and HK Institute of Quantum Science \& Technology, The University of Hong Kong, Pokfulam Road, Hong Kong, China}

	\author{C. M. Wang}
	
	\affiliation{Department of Physics, Shanghai Normal University, Shanghai 200234, China}

	\author{Hai-Zhou Lu}
	\email{Corresponding author: luhaizhou@gmail.com}
	
	\affiliation{Department of Physics, Southern University of Science and Technology (SUSTech),
		Shenzhen 518055, China}
	\affiliation{Quantum Science Center of Guangdong-Hong Kong-Macao Greater Bay Area (Guangdong), Shenzhen 518045, China}
	
	\author{X. C. Xie}
	\affiliation{International Center for Quantum Materials, School of Physics, Peking University, Beijing100871, China}
	\affiliation{Institute for Nanoelectronic Devices and Quantum Computing, Fudan University, Shanghai 200433, China}
	\affiliation{Hefei National Laboratory, Hefei 230088, China}
	
	\begin{abstract}
		A quantized Hall conductance (not conductivity) in three dimensions has been searched for more than 30 years. Here we explore it in 3D topological nodal-ring semimetals, by employing a minimal model describing the essential physics. In particular, the bulk topology can be captured by a momentum-dependent winding number, which confines the drumhead surface states in a specific momentum region. This confinement leads to a surface quantum Hall conductance in a specific energy window in this 3D system. The winding number for the drumhead surface states and Chern number for their quantum Hall effect form a two-fold topological hierarchy.
		We demonstrate the one-to-one correspondence between the momentum-dependent winding number and wavefunction of the drumhead surface states.
		More importantly, we stress that breaking chiral symmetry is necessary for the quantum Hall effect of the drumhead surface states. The analytic theory can be verified numerically by the Kubo formula for the Hall conductance. We propose an experimental setup to distinguish the surface and bulk quantum Hall effects. The theory will be useful for ongoing explorations on nodal-ring semimetals.
	\end{abstract}
	
	\maketitle
\begin{figure}[t!]
		\includegraphics[width=1.0\columnwidth]{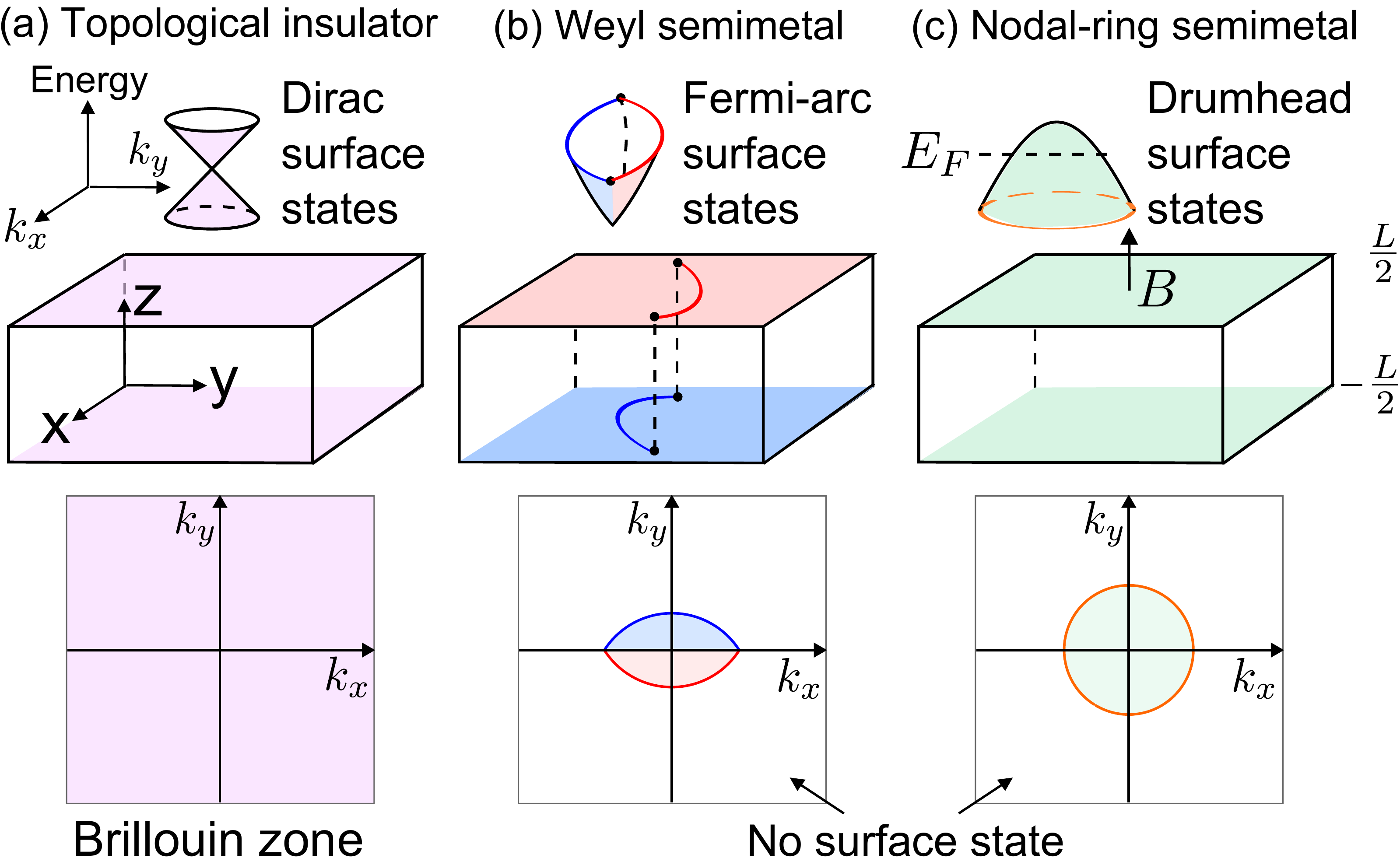}
		\caption{The topologically-protected surface states in topological materials.
			They can have the quantum Hall effect when crossing the Fermi energy $E_F$.
			Mathematically (although not always in the experiments), the surface states of 3D topological insulators can spread over the entire Brillouin zone. By contrast, the surface states in Weyl and nodal-ring semimetals are confined in certain momentum regions [olive and circle area in (b) and (c)], because of topological constrains defined by a $k_z$-dependent Chern number in Weyl semimetals and a $(k_x,k_y)$-dependent winding number in nodal-ring semimetals, respectively.
			Because the drumhead surface states can survive only inside the nodal ring, there is an energy cutoff for the surface quantum Hall effect of the nodal-ring semimetal. By contrast, the energy cutoff in 3D topological insulators is determined by the band edges of the 3D bulk states.
		}\label{Fig:surface}
	\end{figure}

	{\color{blue}\emph{Introduction}.}-
	The quantum Hall effect discovered in 2D electron gas ignites the field of topological phases of matter \cite{Klitzing80prl}. It is characterized by a quantized Hall conductance in units of $e^2/h$ and vanishing longitudinal conductance. In the past over 30 years, much effort has been devoted to realize the quantized Hall conductance in 3D (the 3D quantum Hall effect) \cite{Halperin87jjap,Montambaux90prb,Kohmoto92prb,Koshino01prl,Bernevig07prl,Stormer86prl,Cooper89prl,Hannahs89prl,Hill98prb,Cao12prl,Masuda16sa,Liu16nc, LiuJY19arXiv,Tang19nat,Qin20prl,Zhao21prl-a}. With the discovery of 3D topological materials, a new direction is to use their 2D surface states to host a quantized Hall conductance in 3D, such as in topological insulators \cite{Xu14np,Yoshimi15nc,ZhangSB15srep,Pertsova16prbrc} [Fig.~\ref{Fig:surface}~(a)] and Weyl semimetals [Fig.~\ref{Fig:surface}~(b)]. In Weyl semimetals \cite{Wan11prb,Yang11prb,Burkov11prl,Xu11prl,Delplace12epl,Jiang12pra,Young12prl,Wang12prb,Singh12prb,Wang13prb,LiuJP14prb,Bulmash14prbrc}, the Weyl orbit \cite{Potter14nc,Moll16nat} formed by the Fermi-arc surface states can support a 3D quantum Hall effect with a quantized Hall conductance \cite{WangCM17prl,ZhangC17nc-QHE,Uchida17nc,Schumann18prl,ZhangC19nat,ChenR21prl}.

	In 3D topological nodal-ring semimetals, the conduction and valence energy bands touch to form 1D nodal rings  \cite{Burkov11prb,Chiu14prb,Fang16cpb,Yang17prb,Rui18prbrc,Chen15nl,Bzduvsek16nat,Weng15prb,Yu15prl,Kim15prl,Fang15prbrc,Zhao16prb,Xu17prb,Zhu16prb,Liang16prb,Huang16prbrc,Li16prl,WangJT16prl,SunY17prb,Neupane16prbrc,ChenC17prb,Zeng15arXiv,Chen15nc,Hirayama17nc,Xie15aplm}. They have the topologically protected drumhead surface states on surfaces parallel with nodal rings. It is natural to ask whether the drumhead surface states can host a quantum Hall effect. This topic is addressed by using a non-Hermitian bulk theory, where chiral symmetry is demanded \cite{molina18prl}. The work is well appreciated for the non-Hermitian physics. However, a Hamiltonian for fermions in these materials is Hermitian by nature. Furthermore, chiral symmetry forces a flat surface spectrum, so electrons cannot perform cyclotron motions to host the Landau levels and quantum Hall effect. Therefore, the question remains largely unanswered.

	In this work, we study the quantum Hall effect in topological nodal-ring semimetals, by using a minimal model Hamiltonian that breaks chiral symmetry but carries the essential topological properties, including the winding number and drumhead surface states. The model has a nodal ring lying in the $k_x$-$k_y$ plane. The winding number is topologically nontrivial only within a $(k_x,k_y)$ region confined by the nodal ring, corresponding to topologically protected drumhead surface states on the $z$-direction surfaces [Fig.~\ref{Fig:surface}~(c)]. We analytically relate the $(k_x,k_y)$-dependent winding number to the wavefunction of the surface states. Based on the theory, we show that the drumhead surface states can host a quantized Hall conductance.
	The magnitude and period of the surface quantum Hall conductance can be analytically analyzed by using the Onsager relation and verified by numerical calculations using the Kubo formula for the Hall conductance (Chern number). The winding number for the drumhead surface states and Chern number for their quantum Hall effect form a two-fold topological hierarchy. More importantly, we show that chiral symmetry breaking is essential for the surface Hall effect in topological nodal-ring semimetals.
	The period of the quantized Hall conductance against magnetic field depends on the chiral-symmetry-breaking term in the Hamiltonian. With increasing sample thickness, while the bulk Hall conductance changes drastically, the surface Hall conductance stays robust due to topological protection. Based on this difference, we propose to distinguish surface and bulk Hall conductance in a wedgy sample. The theory will be helpful for ongoing and future experiments in nodal-ring semimetals.

	\begin{figure}[t!]
		\includegraphics[width=1.0\columnwidth]{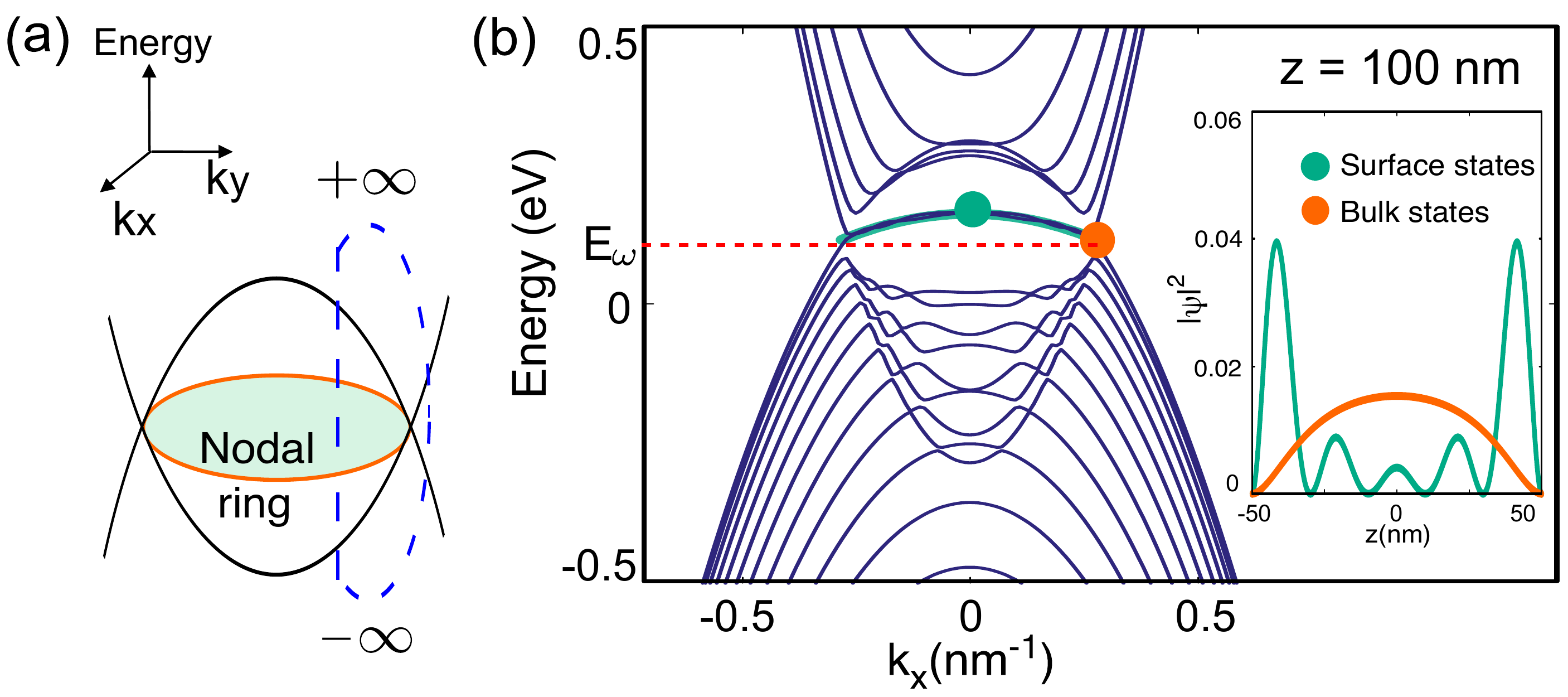}
		\caption{(a) In topological nodal-ring semimetals, the touching points of the conduction and valence bands (black) form a nodal ring (orange ring). An integral of $k_z$ from $-\infty$ to $\infty$ can define a $(k_x,k_y)$-dependent winding number. Within the nodal ring, the winding number is nontrivial, which protects the drumhead surface states on the $z$-direction surfaces [green area in (a) and (b)]. (b) Numerical (dark blue) and analytical [green, Eq.~\eqref{surfacestates}] energy spectrum of a topological nodal-ring semimetal in a slab geometry. Red dashed line marks the energy $E_\omega = C_\bot M_0/M_\bot$. Inset: The wavefunction profiles for typical drumhead surface and bulk states (green and orange dots), respectively.We fit the parameters from the CaAgAs ARPES as $C_\bot$ = $154.0$ eV $\text{\AA}^2$, $C_z$ = $650.0$ eV $\text{\AA}^2$, $M_0$ = $-0.30$ eV, $M_\bot$ = $-386.0$ eV $\text{\AA}^2$, $M_z$ = $-1170.0$ eV $\text{\AA}^2$, and $A = 8.0$ eV $\text{\AA}$. With a weak spin-orbit coupling, as shown in the ARPES \cite{WangXB17prbrc}, the 4-band model of CaAgAs reduces to two degenerate copies of Eq. (\ref{hamiltonian}).}\label{fg:figure1}
	\end{figure}

	{\color{blue}\emph{Minimal model}.}-
	In topological nodal-ring semimetals, the physics near the nodal ring can be essentially captured by a minimal two-band model (see Sec. SVI of \cite{Supp} for a comparison of models for nodal-ring semimetals ),
	\begin{eqnarray}
		\label{hamiltonian}
		\mathcal{H}(\bm{k}) =C_{\bm{k}}+(M_0-M_\bot \mathbf{k}_\parallel^2-M_zk_z^2)\sigma_z+A k_z\sigma_x,
	\end{eqnarray}
	where $\sigma_{x,z}$ are Pauli matrices denoting the orbital degrees of freedom and $\mathbf{k}_\parallel=(k_x,k_y)$. Here $C_{\bm{k}}=C_\bot \mathbf{k}_\parallel^2+C_zk_z^2$ breaks the chiral symmetry $\mathcal{S}=\sigma_y$. The parameters $A$, $M_i$, and $C_i$ can be determined through first-principle calculations or experiments. The seemingly simple model Hamiltonian in Eq.~\eqref{hamiltonian} actually describes a number of realistic materials, such as Ca$_3$P$_2$ \cite{Chan16prb}, TlTaSe$_2$ \cite{Bian16prbrc}, and CaAgX (X= P, As) \cite{WangXB17prbrc,Okamoto16jpsj,Yamakage15jpsj,Takane18npjqm,Xu18prbrc}. The energy spectra of this Hamiltonian read $E_{\pm}=C_{\bm{k}}\pm \sqrt{M_{\bm{k}}^2+(Ak_z)^2}$,  with $M_{\bm{k}}=M_0-M_\bot \mathbf{k}_\parallel^2-M_zk_z^2$. For $M_0M_\bot > 0$, the band-crossing points form a nodal ring, which is located at the $k_z = 0$ plane satisfying $k_x^2+k_y^2 = M_0/M_\bot$ with the energy $E_\omega = C_\bot M_0/M_\bot$.
	
	While our model \eqref{hamiltonian} describes type I nodal-ring semimetals, it can be extended, by adding a tilting term $t(\bm{k})$ to $\mathcal{H}(\bm{k})$ so that $\mathcal{H}^\prime(\bm{k})=\mathcal{H}(\bm{k})+t(\bm{k})\sigma_0$ can describe the type II and III nodal-ring semimetals. It shall be noted that as the tilting term is commonly linear in momentum $\bm{k}$, it will not affect the formation of the 3D Hall effect due to the square term in the surface Hamiltonian of Eq.~\eqref{surfacestates}.
	
	{\color{blue}\emph{Correspondence between $(k_x,k_y)$-dependent winding number and $z$-direction drumhead surface states}.}-
	The model Hamiltonian in Eq.~\eqref{hamiltonian} carries the essential topological properties of topological nodal-ring semimetals \cite{Supp}, including the winding number and drumhead surface states.
	It is well known that the nodal ring is characterized by a quantized $\pi$ Berry phase \cite{Bian16nc}. It is less known the Berry phase is equivalent to the difference between the winding number inside and outside the nodal ring, as shown below. The Berry phase can be obtained by an integral along a closed loop that intersects the nodal ring, as shown by the dashed loop in Fig.~\ref{fg:figure1} (a). Equivalently, the loop can deform to contour the infinity. A winding number $n_w$ can be defined as an integral from $k_z =-\infty$ to $+\infty$ for a given point on the $k_x$-$k_y$ plane (Sec. SI in \cite{Supp}), \begin{eqnarray}
		\label{windingnumber}
		n_w(k_x,k_y) &=& \left\{
		\begin{array}{cc}
			1, & \mathrm{for}\ k_x^2+k_y^2<M_0/M_\perp; \\
			\\
			0, & \mathrm{for}\ k_x^2+k_y^2>M_0/M_\perp, \\
		\end{array}
		\right.
	\end{eqnarray}
	which means that the winding number is nontrivial inside and trivial outside the nodal ring. The difference between the ($k_x,k_y$)-dependent winding number inside and outside the nodal ring is equivalent to the $\pi$ Berry phase that protects the nodal ring. Due to bulk-boundary correspondence, the non-trivial bulk topology property leads to topological surface states at the boundary. Only the winding number inside the nodal ring is nontrivial [green area in Fig.~\ref{fg:figure1} (a)], leading to the formation of drumhead surface states at the boundary along the $z$ direction, as shown in Fig.~\ref{fg:figure1} (b). Note that the ($k_x,k_y$)-dependent winding number has the bulk-boundary correspondence so it can determine where the surface states are in the Brillouin zone, but the $\pi$ Berry phase alone cannot because it is just the difference between the trivial and nontrivial winding number.

	This topological constraint can also be found from directly deriving the wavefunction of the surface states at an open boundary. Taking open boundary conditions in the $z$ direction and keeping periodic boundary conditions in other directions, we can obtain the surface states by adopting the wavefunction ansatz $\Phi=e^{i(k_xx+k_yy)}e^{\lambda z} (\alpha,\beta)^T$, where $T$ is the matrix transposition. After projection, the effective boundary Hamiltonian for the drumhead surface states is obtained as \cite{Supp},
	
	\begin{eqnarray}
		\label{surfacestates}
		H_{\text{surf}} =C_z M_0/M_z+(C_\bot-C_zM_\bot/M_z)\mathbf{k}_\parallel^2,
	\end{eqnarray}
	for $k_x^2+k_y^2<M_0/M_\bot$. The constraint on $k_x,k_y$ means the surface states are bounded by the nodal ring, in accordance with the bulk non-trivial topological invariants.
	
	The above analytical results can be used to identify surface and bulk states. In Fig.~\ref{fg:figure1} (b), we calculate numerically the energy spectrum in a slab geometry, including both surface and bulk modes, as shown by the dark blue lines. From our analytical analysis, the topological surface states are located inside the nodal ring, described by Eq.~\eqref{surfacestates}. For the surface states, the numerical results agree well with the analytical results (green line), allowing us to distinguish them from the bulk modes. Also, the difference between bulk and topological surface states can be viewed in terms of the profile of wavefunctions. In the inset of Fig.~\ref{fg:figure1} (b), two typical wavefunction profiles are plotted, which show that the surface state (green) is localized near the boundaries along the $z$ direction while the bulk state (orange) is extended.

	\begin{figure}[t!]
		\includegraphics[width=1.0\columnwidth]{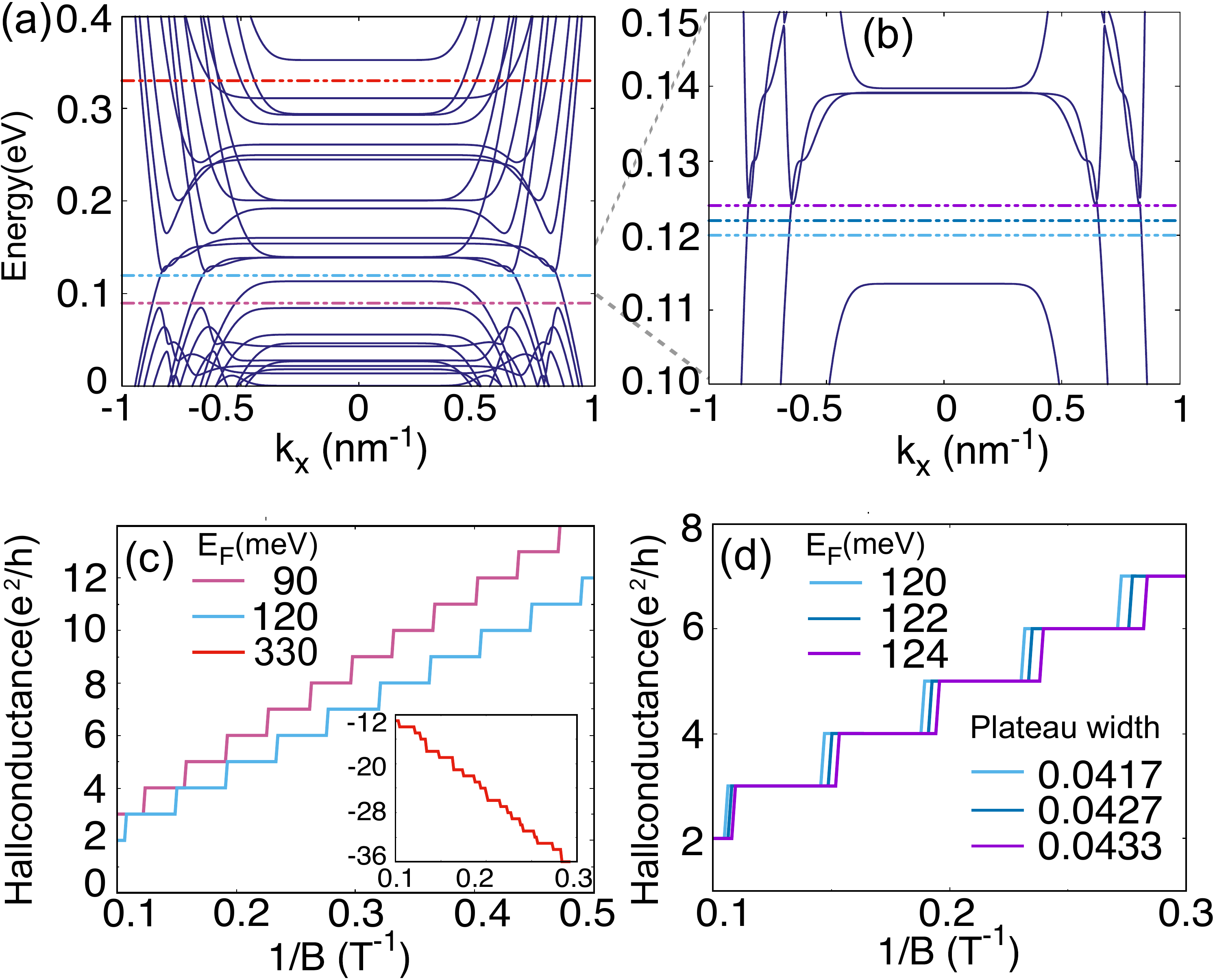}
		\caption{(a) Landau levels of the nodal-ring semimetal slab of thickness $L = 100$ nm, under a $z$-direction magnetic field of $B = 10$ T. $k_x$ is related to the guiding center according to $y_0=k_x\ell_B^2$, with the magnetic length $\ell_B=\sqrt{\hbar/eB}$. Typical energies of $E_\text{F}=$120 meV in the surface region and $E_\text{F}=$(90, 330) meV in the bulk region are shown. (b) Zoom-in of dashed box in (a) shows three typical energies $E_\text{F}=$(120, 122, 124) meV in the surface region. (c) The Hall conductance as a function of $1/B$, corresponding to the energies in (a). (d) The same as (c) except for the energies in (b). The periods of Hall conductance are determined by the width of Hall plateau in unit of T$^{-1}$. The parameters are the same as those in Fig.~\ref{fg:figure1}. }\label{fg:diffe}
	\end{figure}

	{\color{blue}\emph{Two-fold topological hierarchy of the surface Hall conductance.}}-
	The drumhead surface states can contribute to a quantum Hall effect with quantized Hall conductance in the 3D nodal-ring semimetal. We stress that this quantum Hall effect is protected by a two-fold hierarchy. First, the existence of the drumhead surface states is protected by the momentum-dependent winding number, in contrast with 3D topological insulators and Weyl semimetals in Figs.~\ref{Fig:surface}~(a) and (b). Second, the quantum Hall effect of the drumhead surface states is protected by the usual Chern number \cite{Thouless82prl} but uniquely, because the surface quantum Hall effect of the drumhead surface states can only occur in an energy window [Eq.~\eqref{surfacerange} or (\ref{surfacerange-2})] due to the topological confinement imposed by the momentum-dependent winding number, as shown below.
	
	The two-fold topological hierarchy imparts a unique energy and thickness dependence to the quantum Hall effect of the drumhead surface states, effectively distinguishing them from the bulk states. This distinction is of significance for identifying nodal-ring semimetals in quantum transport experiments.

	{\color{blue}\emph{Significance of breaking chiral symmetry in surface Hall response}.}-
	In order to form Landau levels and quantum Hall effect from surface states, the quadratic $C_{\bm k}$ terms in Eq.~\eqref{surfacestates} are necessary. Without the $C_{\bm k}$ terms and the dispersion they create, the surface electrons have no velocity thus cannot perform the cyclotron motion in response to a magnetic field and cannot support Landau levels and quantum Hall effect. It is noteworthy that the $C_{\bm{k}}$ term breaks the chiral symmetry, making the topological surface modes dispersive and away from zero energy. This is in sharp contrast with the earlier work~\cite{molina18prl}, where the dispersion term $C_{\bm{k}}$ must be constant due to chiral symmetry, resulting in the absence of Landau levels from the surface states.
	
	In Fig.~\ref{fg:diffe} (a), we plot the Landau levels for the topological nodal-ring semimetal in the slab geometry under an applied $z$-directional magnetic field with strength $B=10$ T. The Landau levels are functions of $k_x$, because when considering the edges along the $y$ direction, the Landau levels deform with the guiding center $y_0\equiv k_x\ell_B^2$, with the magnetic length $\ell_B\equiv \sqrt{\hbar/eB}$. The conduction and valence bands are not symmetric due to the breaking of chiral symmetry. We focus on the Landau levels from the drumhead surface states.
	Due to its topological origin, the surface states are confined in the region $0<k_x^2+k_y^2<M_0/M_\bot$, distinct from usual 2D electron gas.Applying this constraint to Eq.~\eqref{surfacestates}, we can get the energy window of the surface states
	\begin{eqnarray}
		\label{surfacerange}
		C_\bot M_0/M_\bot<E_\text{F}<C_z M_0/M_z,
	\end{eqnarray}
	if $C_\bot<C_zM_\bot/M_z$ (like that in the material CaAgAs),
	or
	\begin{eqnarray}
		\label{surfacerange-2}
		C_\bot M_0/M_\bot>E_\text{F}>C_z M_0/M_z,
	\end{eqnarray}
	if $C_\bot>C_zM_\bot/M_z$. This energy window of the surface states is material-dependent. In this region, the surface spectrum at a given Fermi energy can form a complete Fermi circle, which means that electrons can undergo a complete cyclotron motion in a perpendicular magnetic
	field to form Landau levels. It should be emphasized here that the energy cutoff in Eq.~\eqref{surfacerange} or \eqref{surfacerange-2} is enforced by the momentum-dependent winding number of Eq.~\eqref{windingnumber}. Thus, the energy cutoff is a direct manifestation of the two-fold hierarchy of topological protection.

	Next we apply the standard Kubo formula to calculate the Hall conductivity (Sec. SIII of \cite{Supp}),
	\begin{eqnarray}\label{kubo}
		\sigma_{xy}=\frac{e^2\hbar}{iV}\sum_{\alpha,\beta(\ne \alpha)}\frac{\langle u_\alpha |v_x|u_\beta\rangle\langle u_\beta |v_y|u_\alpha\rangle}{(E_\alpha-E_\beta)^2}[f(E_\alpha)-f(E_\beta)], \nonumber
	\end{eqnarray}
	where $|u_\alpha\rangle$ is the eigenstate of energy $E_\alpha$ for the slab in a $z$-direction magnetic field and with open boundaries at $z = \pm L/2$, $v_x$ and $v_y$ are the velocity operators, $f(x)$ is the Fermi distribution, $V$ is the volume of the slab. By tuning the Fermi energy to the region in Eq.~\eqref{surfacerange}, we can calculate the surface conductance, as shown in Fig.~\ref{fg:diffe} (d). $\sigma_{xy}$ exhibits distinctive integer plateaus as a function of the inverse of magnetic field, showing that the quantum Hall effect arises from drumhead surface states.
	
	Two remarks are in order. First, to observe the quantized Hall conductance of the drumhead surface states, it is required to deplete the bulk carriers by tuning the Fermi energy to the energy at the nodal ring. Second, the transport between the two surfaces of the nodal-ring semimetals is not necessarily for the formation of 3D quantum Hall effect, because the drumhead surface states can form a closed Fermi loop needed by Landau levels and the quantum Hall effect on a single surface. Nevertheless, it is possible for the nodal ring to connect the top and bottom surface states to give a 3D quantum Hall effect.

	{\color{blue} \emph{Nodal-ring semimetal} CaAgAs.}--We calculate
	the Hall conductance using the parameter fitted for an experimentally realized material CaAgAs \cite{WangXB17prbrc}. With a weak spin-orbit coupling, as shown in the ARPES \cite{WangXB17prbrc}, the 4-band model of CaAgAs reduces to two degenerate copies of Eq.~(\ref{hamiltonian}). For the surface states of CaAgAs, the period of the surface Hall conductance under applied magnetic fields can be analytically obtained by the Onsager relation
	$\Delta(1/B) = -(2\pi e/\hbar)(1/A_s)$, where $\Delta$ denotes the period with respect to $1/B$, which is equal to the width of Hall plateau, $A_s$ is the maximum area through which the Fermi energy cuts in $k$ space. With the surface Hamiltonian in Eq.~\eqref{surfacestates}, the Onsager relation can be determined as $\Delta(1/B) = -2e(C_\bot-C_zM_\bot/M_z)/\hbar(E_\text{F}-C_z M_0/M_z)$. As $\Delta(1/B)$ depends positively on $E_\text{F}$, the period of Hall conductance increases when increasing the Fermi energy.
	Figure~\ref{fg:diffe} (d) shows the Hall conductance for three typical Fermi energies in the surface energy window in Fig.~\ref{fg:diffe} (b). We compare the periods obtained from the surface theory and the numerical results determined by the width of Hall plateau in Fig.~\ref{fg:diffe} (d). The consistence between analytical and numerical results further confirms that the Hall conductance at these energies originates from the drumhead surface states (Sec. SIV of \cite{Supp}).
	
	We can further tune the Fermi energy away from the surface energy window and calculate the bulk Hall conductance, as shown in Fig.~\ref{fg:diffe} (c) for one energy in the surface region and two energies in the bulk region in Fig.~\ref{fg:diffe} (a). The Hall conductance is positive (negative) for electron (hole) carriers. Outside the surface energy window, the Hall conductance can no longer be captured by the surface theory, the Hall conductance becomes multi-periodic as more bulk subbands emerge.
	
	\begin{figure}[t!]
		\includegraphics[width=1.0\columnwidth]{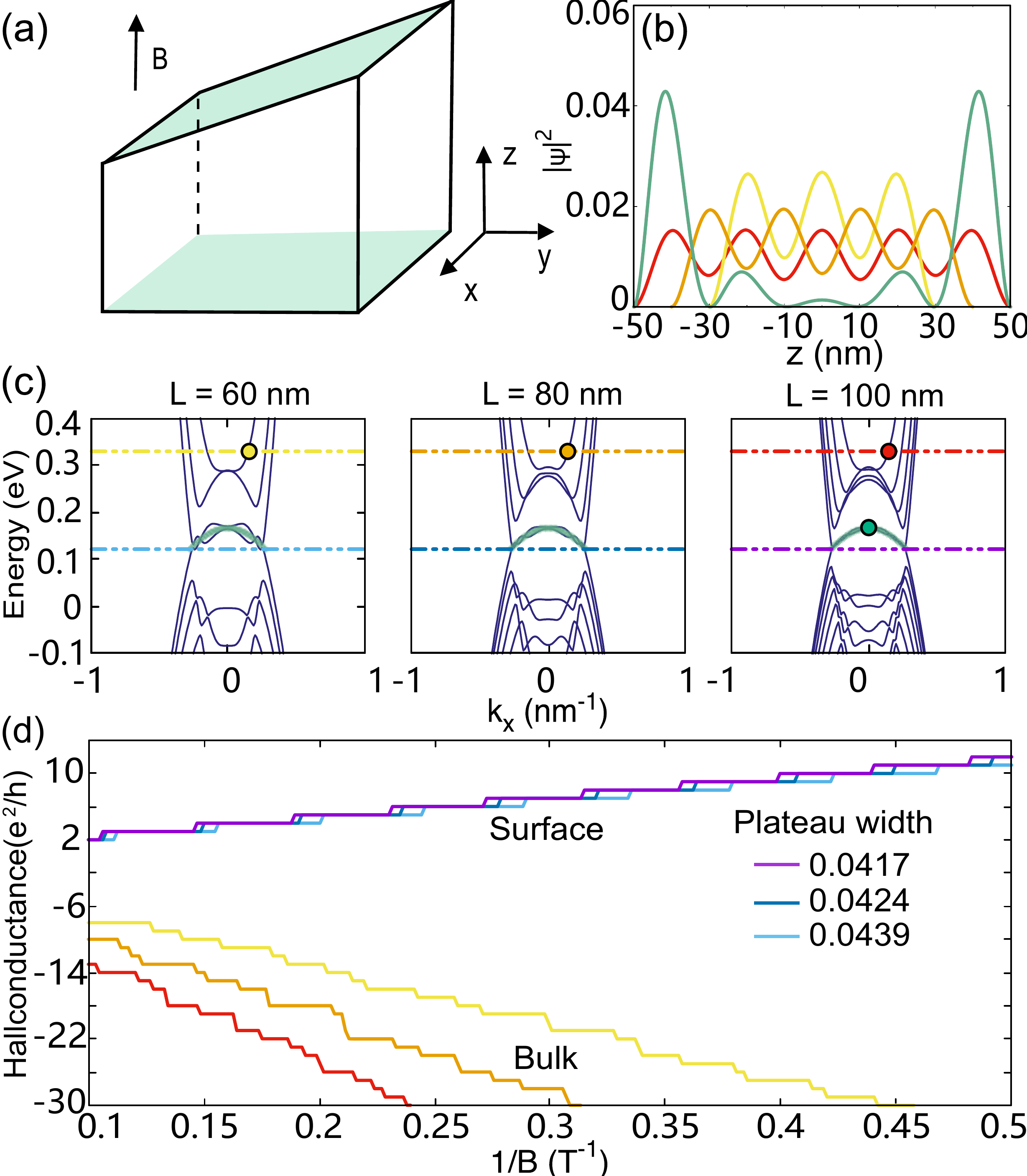}
		\caption{(a) Schematic of the wedge geometry. (b) Wavefunction profiles for the states marked by dots in (c). (c) Energy spectra for the slab with thicknesses of 60, 80, and 100 nm. The lower dashed line denotes the Fermi energy ($E_\text{F} = 0.12$ eV) that cuts through surface states, while the upper dashed line denotes the Fermi energy ($E_\text{F} = 0.33$ eV) that cuts through bulk states. The green curves denote the analytical result of the surface-state spectrum. (d) The Hall conductance for Fermi energies at surface (upper three) and bulk (lower three) energy regions. The colors of curves correspond to the dashed lines in (c).The parameters are the same as those in Fig.~\ref{fg:figure1}.}\label{fg:diffl}
	\end{figure}

	{\color{blue}\emph{Distinguish surface and bulk contributions}.}-
	By noticing that the sample thickness affects surface and bulk states differently, we show that two kinds of Hall conductance can be identified by varying the sample thickness in a wedgy device, as shown in Fig.~\ref{fg:diffl} (a). The wedgy device has been practiced in experiments to control sample thickness by varying growth time \cite{Moll16nat,ZhangC19nat}. We compare three thicknesses 60, 80, and 100 nm. We focus on two Fermi energies, one in the surface energy window and the other in the bulk energy region, as shown by the dashed lines in Fig.~\ref{fg:diffl} (c). By varying the sample thickness, the surface spectrum (green) is basically unchanged, while more subbands (dark blue) emerge in the bulk spectrum. In Fig.~\ref{fg:diffl} (b), the wavefunctions show that the surface states are always localized at the boundary while the extension of bulk states changes between different subbands.
	
	Figure~\ref{fg:diffl} (d) shows the Hall conductance for the three thicknesses, with the colors corresponding to the dashed lines in Fig.~\ref{fg:diffl} (c). In Fig.~\ref{fg:diffl}(c), the upper three curves are calculated with $E_F$ in the energy range of the surface states, so they are denoted as the surface Hall conductance; the lower three curves are calculated with $E_F$ in the energy region of the bulk states, so they are denoted as the bulk Hall conductance. For the surface quantum Hall conductance, its magnitude is positive and remains basically unchanged for different thicknesses. By increasing thickness, the period of the surface Hall conductance gradually approaches the theoretical value of $0.0392$ T$^{-1}$ determined by the Onsager relation. In contrast, for the bulk conductance at a fixed Fermi energy, with increasing sample thickness, both the magnitude and period of the conductance change drastically. The difference between surface and bulk quantum Hall conductance is due to the emerging subbands crossing the Fermi energy. By varying the sample thickness continuously, there are no other subbands emerging in the energy range of the surface states, while there are new sub-bands in the range of the bulk states. These subbands host more Landau levels and increase the Hall conductance. The Fermi surfaces of these bulk subbands are different, leading to the multi-periodic quantum Hall conductance, serving as another signature for the bulk states.

	\vspace{0.5cm}
	
	\noindent
	{\color{blue}\emph{Declarations}}
	
	\noindent
	\textbf{Ethics approval and consent to participate}\\
	Not applicable.
	
	\vspace{0.5cm}
	
	\noindent
	\textbf{Consent for publication}\\
	Not applicable.
	
	\vspace{0.5cm}
	
	\noindent
	\textbf{Availability of data and materials}\\
	All data underlying the results are available from the authors upon reasonable request.
	
	\vspace{0.5cm}
	
	\noindent
	\textbf{Competing interests}\\
	The authors declare no competing interests.
	
	\vspace{0.5cm}
	
	\noindent
	\textbf{Funding}\\
	This work was supported by the National Key R$\&$D Program of China (2022YFA1403700), the Innovation Program for Quantum Science and Technology (2021ZD0302400), the National Natural Science Foundation of China (11534001, 11974249, and 11925402), Guangdong province (2020KCXTD001 and 2016ZT06D348), the Strategic Priority Research Program of Chinese Academy of Sciences (Grant No. XDB28000000), the Natural Science Foundation of Shanghai (Grant No. 19ZR1437300), the Shenzhen High-level Special Fund (No. G02206304, G02206404), and the Science, Technology and Innovation Commission of Shenzhen Municipality (ZDSYS20170303165926217, JCYJ20170412152620376, KYTDPT20181011104202253). The numerical calculations were supported by Center for Computational Science and Engineering of SUSTech.
	
	\vspace{0.5cm}
	
	\noindent
	\textbf{Authors' contributions}\\
	GZ, SL, and WR contribute to the work equally. HL and XX supervised the work. GZ and WR did the theoretical analysis with the input from CW, HL, and XX. GZ and SL performed the numerical calculations, with the input from CW, HL, and XX. All authors participate in the analysis of the results. GZ, WR, and HL wrote the manuscript with the input from all authors. All authors have read and approved the manuscript.
	\vspace{0.5cm}
	
	\noindent
	\textbf{Acknowledgements}\\
	Not applicable.

	\bibliographystyle{apsrev4-1-etal-title}
	\bibliography{refs-transport}

\end{document}